\documentclass[aps,prd,twocolumn,showkeys,superscriptaddress, 10pt]{revtex4-2}
\usepackage{amssymb}
\usepackage{tikz}
\usepackage{graphicx}
\usepackage{dcolumn}
\usepackage{bm}
\usepackage{relsize}
\usepackage{lipsum}
\usepackage[margin=0.7in]{geometry}
\usepackage{amsmath,amssymb}    
\usepackage{url}
\newcommand{\beq}{\begin{equation}}
\newcommand{\eeq}{\end{equation}}

\begin{document}



\title{Optimal Design of Coatings for Mirrors of Gravitational Wave Detectors: Analytic  \emph{Turbo} 
Solution via Herpin Equivalent~Layers}


\author{Vincenzo Pierro}
\affiliation{Dipartimento di Ingegneria (DING),  Universit\'{a} del Sannio, I-82100 Benevento, Italy.}
\affiliation{INFN, Sezione di Napoli Gruppo Collegato di Salerno,  Complesso Universitario di Monte S. Angelo, I-80126 Napoli, Italy.}
\email[]{corresponding: pierro@unisannio.it, vpierro@sa.infn.it}

\author{Vincenzo Fiumara}
\affiliation{Scuola di Ingegneria, Universit\'{a} della Basilicata, I-85100 Potenza, Italy.}
\affiliation{INFN, Sezione di Napoli Gruppo Collegato di Salerno,  Complesso Universitario di Monte S. Angelo, I-80126 Napoli, Italy.}

\author{Francesco Chiadini}
\affiliation{Dipartimento di Ingegneria Industriale, DIIN, Universitá di Salerno, I-84084 Fisciano, Salerno, Italy}
\affiliation{INFN, Sezione di Napoli Gruppo Collegato di Salerno,  Complesso Universitario di Monte S. Angelo, I-80126 Napoli, Italy.}

\begin{abstract}
In this paper, an analytical solution to the problem of optimal dielectric coating design of mirrors for gravitational wave detectors is found.
The technique used to solve this problem is based on Herpin's equivalent layers, which provide a simple, constructive, and analytical solution.
The performance of the Herpin-type design exceeds that of the periodic design and is almost equal to the performance of the numerical, non-constructive optimized design obtained by  
\emph{brute force.} 
Note that the existence of explicit analytic constructive solutions of a constrained optimization problem is not guaranteed in general,
 when such a solution is found, we speak of \emph{turbo} optimal solutions.
\end{abstract}

\vspace{2pc}
\keywords{Dielectric multilayers; Gravitational Wave Interferometric Detectors; Coating Optimization}

\maketitle

\section{Introduction}
\label{S:1}
{The development of optimized coatings for the end test-masses of the gravitational wave interferometers is one of the important goals to be achieved for improving the sensitivities of gravitational wave detectors} \cite{Virgo,LIGO}. 
{ Indeed, the~coating Brownian noise is the most relevant source of noise in the band of interest for astrophysical observation.
To reduce this source of noise, researchers can act by using different materials with the best properties for the mirrors, optimizing the interferometer's laser beam, 
lowering the temperature to cryogenic values, and~finally acting on the coating design. 
Unfortunately, there are not many glassy materials that satisfy the optical requirements necessary for gravitational wave detectors, 
the improvement of the laser beam would require a rethinking of the whole interferometer cavity as well as make this cavity cryogenic 
(see the papers~\cite{adLIGO, adVirgo, KAGRA} for a review on the subject).

In this work, we explore the possibility of optimizing the design of the mirrors by acting on the thicknesses of the layers that form the coating, with~the materials currently available.}
Although coatings made of multiple materials~\cite{Yam, Stein, PRRvp} (obtained, in~some cases, from~cascades of binary designs) 
have been recently proposed, 
further study of binary coating theory provides the theoretical tools to understand more complex~approaches.

\begin{figure}[h]
 \includegraphics[width=0.98\linewidth]{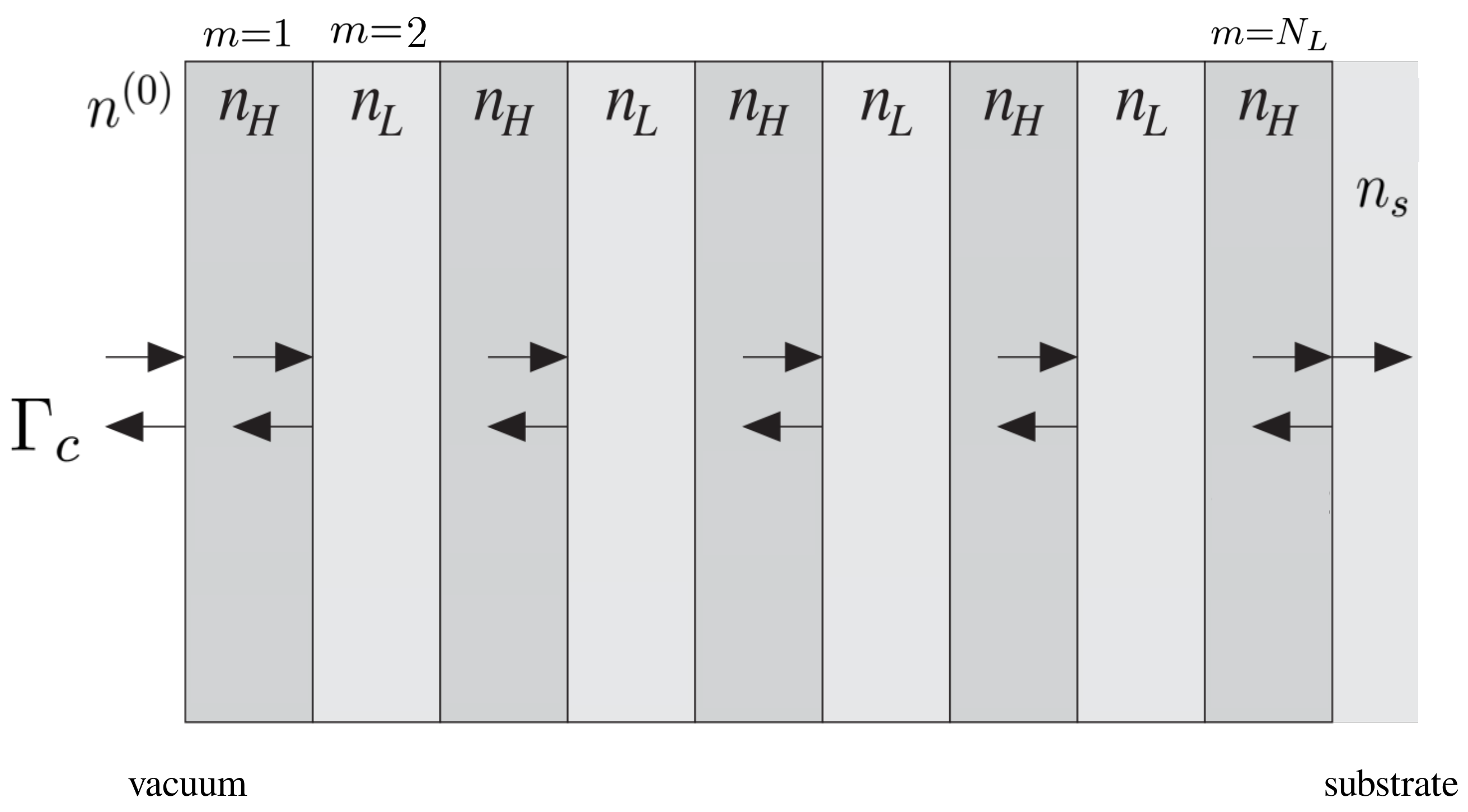}
\caption{In 
 the figure an example of binary coating made of alternating $L$ and $H$ type materials is displayed. The~laser light 
is normally incident form the left (i.e., a monochromatic wave, an~$\exp \imath \omega t$ dependence on time is supposed), 
$\Gamma_c$ is the reflection coefficient of the stratified medium. 
The refractive index of the $m$-th layer is in general $n^{(m)}$, for~binary coating the sequence of refractive index is 
alternating with values $n_H$ (high refractive index) and $n_L$ (low refractive index) . The~substrate refractive index is $n_s$, the~leftmost half space 
has, refractive index $n^{(0)}$ (here it is supposed to be the vacuum).}
\label{fig1}
\end{figure}

It is well known~\cite{Rayleigh} that the electrodynamics of multilayers structures, like those depicted in Figure~\ref{fig1}, can be described in a semi-analytic way with the method of the characteristic 
matrices of the layers (also called transmission matrices~\cite{Abeles}).
The characteristic matrix of the $m$-th layer can be written~\cite{BornWolf, Orfanidis}:
\begin{equation}
\mathbf{T}_m=
\begin{bmatrix}\tabcolsep6pt
\cos\left(\psi_m \right)&  \imath \displaystyle  (n^{(m)})^{-1} \sin\left(\psi_m\right)\\[10pt]
\displaystyle \imath n^{(m)} \sin\left(\psi_m\right)&
\cos\left(\psi_m\right)
\end{bmatrix},
\label{Tmatrix}
\end{equation}
where $\psi_m=\frac{2 \pi}{\lambda_0}n^{(m)}d_m
$
is the electric phase, $\lambda_0$ and $d_m$ are the free space wavelength and the layer thickness, respectively, and~$n^{(m)}= n^{(m)}_r - \imath \kappa^{(m)}$ is the complex refractive index.
Here  $\kappa^{(m)}$ is the extinction coefficient, which for the considered materials will be negligible (i.e., $\kappa^{(m)}\sim 10^{-8}$).

\vspace{-6pt}

The optical response of the whole coating {  (i.e., the transmittance)} can be computed from the multilayer characteristic matrix,
\beq
\mathbf{T}=\mathbf{T}_1 \cdot \mathbf{T}_1 \cdot ... \cdot \mathbf{T}_{N_L}
\label{charmat}
\eeq
where $N_L$ is the total number of layers numbered from the vacuum to the substrate as illustrated in Figure~\ref{fig1}.
To write Equation~(2) we use the property that the characteristic matrix of a sequence of layers is the product of the characteristic matrices of the individual~layers.

{ The transmittance calculation is done in two steps, first the equivalent reflection index $n_c$
of the multi-layer structure, and~then the reflection coefficient $\Gamma_c$ at the vacuum interface are computed.}
The complex reflection coefficient $\Gamma_c$ at the vacuum/coating interface is given by: 
\beq
\Gamma_c=\frac{n^{(0)}-n_c}{n^{(0)}+n_c}
\eeq
where $n_c$ is the effective refractive index of the whole multilayer structure,
\beq
n_c=\frac{T_{21}+n_s T_{22}}{T_{11}+n_s T_{12}}.
\eeq
The 
 power transmittance at the vacuum/coating interface is $\tau_c=1 - |\Gamma_c|^2$.
In the case of binary coating we have: 
\beq
\left \{
\begin{array}{ll}
n^{(m)}= n_H - \imath \kappa_H, \mbox{ $m$ odd}, & n^{(m)}= n_L - \imath \kappa_L, \mbox{ $m$ even};\\ 
\phi_m = \phi_H, \mbox{$m$ odd }, & \phi_m = \phi_L, \mbox{ $m$ even} ;\\ 
Y_m = Y_H, \mbox{ $m$ odd}, & Y_m = Y_L, \mbox{$m$ even }.
\end{array}
\right.
\eeq 

These 
multilayer structures are made of $N_L$ alternating high $n_H$ and low $n_L$ refractive indexes 
deposited on a substrate of refractive index $n_s$. The~coefficients $\phi_H$ and $\phi_L$ are the mechanical losses and $Y_H$ and
$Y_L$ are the Young moduli of the two materials. Let us also introduce the specific noise coefficients that will be used in the following:
\beq
\eta_{L}= \frac{1}{\sqrt{\pi} w}  \phi_{L} \left( \frac{Y_{L}}{Y_s} + \frac{Y_s}{Y_{L}} \right) \,\,\,\,\,
\eta_{H}= \frac{1}{\sqrt{\pi} w}  \phi_{H} \left( \frac{Y_{H}}{Y_s} + \frac{Y_s}{Y_{H}} \right),
\eeq
where $w$ is the (assumed Gaussian) laser-beam waist.
Thermal noise in gravitational detectors is the most important limitation for their operation.
We refer the reader to the works in~\cite{Harry, Abernathy18, Flaminio10} for an exhaustive description of the problem and the proposed solutions~\cite{pinard11, pinard17}
in the operating observatory Virgo~\cite{Virgo} and LIGO~\cite{LIGO}.

A straightforward  formulation of the coating optimization problem for the design of low noise dielectric mirror can be:
\begin{equation}
\begin{aligned}
& \underset{z_1,..., z_{N_L} \in \Omega}{\text{Minimize}}
& & \bar{\phi}_c \\
& \text{subject to}
& & \tau_c \leq \tau_0 \;
\end{aligned}
\label{optprob}
\end{equation}
where the constraint transmittance $\tau_0$ should be a few parts per million (ppm), typically $\tau_0 = 6$ ppm.

Defining the normalized loss angle $\bar{\phi}_c= \phi_c /(\lambda_0 \eta_L)$ and introducing
the normalized physical length $z_m= d_m/\lambda_0$, 
where $\lambda_0$ is the free-space wavelength of the laser, we have for  binary 
coatings:
\beq
\bar{\phi}_c=    {\,}^{o}\mathlarger{\Sigma}_{m=1}^{N_L} \gamma z_{m}  +  {\,}^{e}\mathlarger{\Sigma}_{m=1}^{N_L} z_{m},
\label{normirnoise}
\eeq
where summation $^{e}\mathlarger{\Sigma}$ (resp. $^{o}\mathlarger{\Sigma}$) is on the  even (resp. odd) integer such that $1 \le m \le N_L$.
The noise ratio coefficient in Equation~(\ref{normirnoise}), i.e.,~$\gamma=\eta_H/\eta_L$,  can be explicitly written for binary coating as:
\beq
\gamma = \frac{\phi_{H}}{\phi_{L}}\left( \frac{Y_{H}}{Y_s} + \frac{Y_s}{Y_{H}} \right) \left( \frac{Y_{L}}{Y_s} + \frac{Y_s}{Y_{L}} \right)^{-1}.
\label{gamma}
\eeq

In the case where the refractive index $n_L$ is the same as that $n_s$ of the substrate material (as in current gravitational wave detectors) 
$N_L$ is an odd number.
{The choice of an even $N_L$  would lead to a configuration with the rightmost layer (near the substrate) 
made of low refractive index material (the same as the substrate) that would increase noise without having any effect on reflectivity.}
The search space $\Omega$ is defined by the inequalities $0 \le z_m \le 0.25/n_H$ for  odd $m$ and  $0  \le z_m \le 0.5/n_L$ for  even $m$.
An alternative and equivalent way to formulate the optimization problem (as shown in~\cite{OptMat}) is
\begin{equation}
\begin{aligned}
& \underset{z_1,..., z_{N_L} \in \Omega}{\text{Minimize}}
& & \tau_c  \\
& \text{subject to}
& & \bar{\phi}_c \leq \bar{\phi}_0 \; 
\end{aligned}
\label{optprob1}
\end{equation}
where $\bar{\phi}_0$ is a prescribed maximum allowed loss~angle.

\section{The Herpin Equivalent Layer Optimization~Problem}
\label{S:2}
According to Herpin's equivalent layer theorem~\cite{Herpin1} a symmetrical multilayer stack (i.e., a~{\it palindrome} sequence of dielectric thin films) 
is equivalent to a single layer.
This theorem is based on the fact that the two elements on the main diagonal of the characteristic matrix of any palindromic sequence of materials are~equal.

In this paper, we consider an equivalent Herpin layer that consists of three physical layers arranged in an $LHL$-type sequence.
According to the general theorem, this sequence must be dielectrically and geometrically symmetric, so both materials and layer thicknesses must be palindromic.
Thus denoting by $p$ and $q$ the normalized lengths of the layers $L$ and $H$ respectively,
below are shown, using a simple computation, the~relevant elements of the transmission matrix
 $\mathbf{T}^{(E)}$ of the considered {\it virtual layer} $E=LHL$:

\begin{widetext}
\beq
T^{(E)}_{11}=
\cos \left(2 \pi  q n_H\right) \cos \left(4 \pi  p n_L\right)\,\,
-\frac{\left(n_H^2+n_L^2\right) \sin \left(2 \pi  q n_H\right) \sin \left(4 \pi  p n_L\right)}{2 n_H n_L},
\eeq
\end{widetext}
\begin{widetext}
\beq
T^{(E)}_{12}=
i \left(-\frac{n_H \sin \left(2 \pi  q n_H\right) \sin ^2\left(2 \pi  p n_L\right)}{n_L^2} 
+ \frac{\sin \left(2 \pi  q n_H\right) \cos ^2\left(2 \pi  p n_L\right)}{n_H}
+\frac{\cos \left(2 \pi  q n_H\right) \sin
   \left(4 \pi  p n_L\right)}{n_L}\right).
\eeq
\end{widetext}

It can be verified by inspection, and this is also the main result of Herpin's theorem,  that $T^{(E)}_{11}=T^{(E)}_{22}$, 
and because of the unitary constraint on the determinant of the characteristic matrix $T$, 
{ the last element is uniquely determined by solving the following equation w.r.t. $T^{(E)}_{21}$:}
\beq
(T^{(E)}_{11})^2-T^{(E)}_{12}T^{(E)}_{21}=1.
\eeq

To understand the reasons that lead to formulating the present analytical solution, we summarize the results of the papers~\cite{OptMat,Spie}.
In the paper~\cite{OptMat}, a~multi-objective optimization (with the BorgMOEA algorithm~\cite{borgMOEA}) without {\it a priori} hypothesis was applied to the problem of general optimization 
of binary coatings for gravitational wave detectors~mirrors.

It has been shown in~\cite{OptMat} that the Pareto front remains the same whether one sets up a code that solves problem (\ref{optprob}) or implements problem (\ref{optprob1}), 
so the two formulations considered are equivalent.
Moreover, in~the same paper we show that the optimal design is made by the following sequences of layers  $L(HL)^{N_D}H$.
The sequence of
thicknesses associated with this solution is given by $z_{L,i}(z_{H}z_{L})^{N_D}z_{H,f}$, where in general $z_{L,i} \ne z_{L}$, $z_{H,f}\ne z_{H}$.
This solution is periodic except for the first and last layers, and~that is why the solution has been called periodic with initial and final tweaking.
Furthermore $z_{L}, \,\, z_H$ satisfy an approximate Bragg condition $n_L z_L + n_H z_H \sim 0.5$.

In the paper~\cite{Spie},  a very simple periodic solution $(HL)^{N_D}H$ was studied  and experimentally validated~\cite{Villar10}. In~these articles the tweaking procedure has been considered but only as a possible second step of improvement of the periodic design, considering the thicknesses of the innermost layers fixed to the values calculated in the first step.
This solution approximates that which would be found by optimizing on all four layers simultaneously, that is implemented in~\cite{OptMat}.


Finally, in~the paper~\cite{OptMat} (see Equation~(20) therein) it has been shown that the Pareto front of the optimized solutions is placed close to the transmittance versus thermal 
noise line relative to a suitable (virtual) quarter-wave~design.

{ So far, we have mentioned almost exhaustively all the existing literature on the coating optimization problem.
For the sake of completeness, although~not completely relevant, let us mention~\cite{Russian} which proposes a physical-mathematical 
approach to the computation of the best periodic design, we emphasize that the method is not based on solving an optimization problem.}


Taking into account the hints introduced above, we assume that the optimal solution is of the form $(EL)^{N_D}E$ where $E$ is an equivalent Herpin layer, as~introduced above, 
and $L$ is a quarter-wave layer. 
{ We note that in these designs the last layer is of type $E$, i.e.,~it is a {\em virtual layer} consisting of an $LHL$ sequence.
Additionally, in this case, the last physical layer (near the substrate) of type $L$ in the last  {\em  virtual layer} $E$ is not 
considered because it is made of the same material as the substrate and does not contribute to the dielectric contrast 
(actually, the last interface does not exist, it is fictitious).}

We are now able, from~all these ansatz introduced above, to~reformulate the optimization problem as follows:
\begin{equation}
\begin{aligned}
& \underset{p,q,N_D}{\text{Minimize}}
& & \bar{\phi}_c \\
& \text{subject to}
& & E \in QWL \,\,\text{and}\,\, \tau_c \leq \tau_0 \;
\end{aligned}
\label{optherp}
\end{equation}
here $QWL$ is the set of quarter wavelength transmission matrix and $\bar{\phi}_c$ is determined by the normalized physical length $p$, and~$q$.
We have 
\beq
\bar{\phi}_c= (N_D + 1)\gamma q + (2 N_D +1) p + N_D p_{1/4},
\label{eq:hernoise}
\eeq
where $p_{1/4}=1/(4 \, n_L)$ is the normalized quarter wavelength thickness of $L$ layer.
The condition $E \in QWL$ can be explicated by requiring $T^{(E)}_{11}=0$ i.e.,
\begin{widetext}
\beq
\cos \left(2 \pi  q n_H\right) \cos \left(4 \pi  p n_L\right) -\frac{\left(n_H^2+n_L^2\right) \sin \left(2 \pi  q n_H\right) \sin \left(4 \pi  p n_L\right)}{2 n_H n_L}=0.
\label{eq:costr}
\eeq
\end{widetext}

\section{Numerical~Results}
\label{S:3}
In this section, we compute the Pareto front of the three competing designs , i.e.,~the proposed design method in Equation~(\ref{optherp}), the~periodic design~\cite{Spie} and the fully ({\it brute force}) optimized design~\cite{OptMat}.
Table~\ref{tab:partab} is used  as a reference for the physical parameters of the $L$ and $H$ materials.

\begin{figure}[h]
\includegraphics[width=0.76\linewidth]{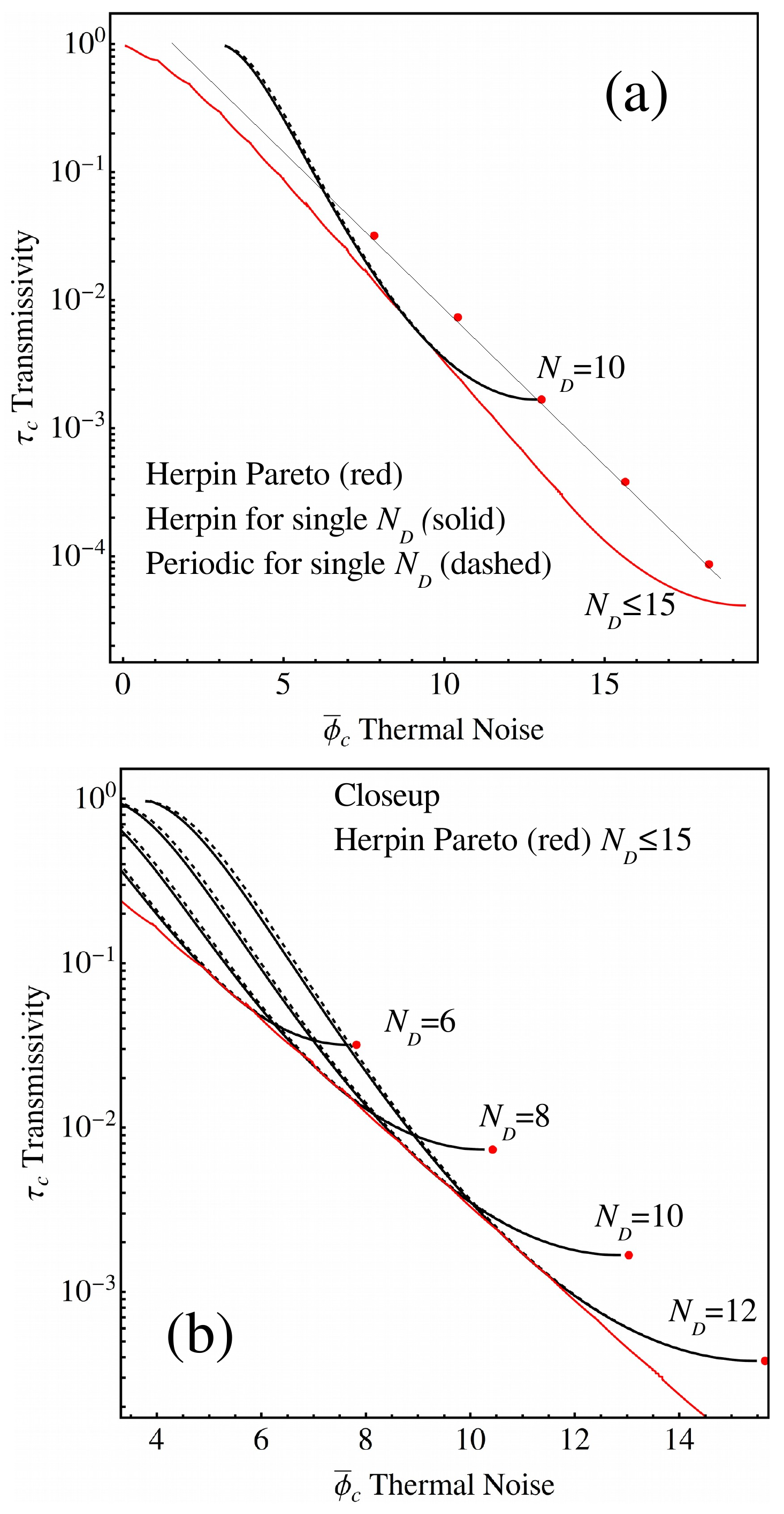}
\caption{The figure shows in (\textbf{a}), in~red, the~Pareto front by Herpin's method ${\mathcal P}_H(\bar{\phi}_c)$ obtained by keeping $N_D \le 15$ { as a 
function of dimensionless normalized mechanical loss $\bar{\phi}_c$}. 
For fixed $N_D=10$ the Herpin Pareto front (solid black) and the periodic Pareto front (dashed black) are displayed for comparison.
Below in (\textbf{b}) a closeup of the central part of (\textbf{a}) where some additional fixed $N_D$ Pareto curves has been displayed. 
The red dots in (\textbf{a},\textbf{b}) are the quarter-wavelength designs for $N_D=6, 8, 10, 12$.}
\label{fig2}
\end{figure}

\begin{table}[t]
\centering
\begin{tabular}{|l| l|}
\hline \hline
Coating  & Substrate \\
$H$ {\small (amorphous Ti-doped Ta$_2$O$_5$ )} &  \small (bulk crystalline SiO$_2$)\\  
$L$ {\small (SiO$_2$)} &    \\  \hline
$n_H = 2.10$  & $n_s=1.45$ \\ \hline
$n_L = 1.45$  &  $Y_s = 72$ GPa  \\ \hline
$\kappa_H = 4.0 \times 10^{-8}$  & $\kappa_s=8.4 \times 10^{-11}$ \\ \hline
$\kappa_L =  8.4 \times 10^{-11}$  & $\phi_s = 7.00 \times 10^{-8}$   \\ \hline
$Y_H = 147$ GPa &   \\ \hline
$Y_L =  72$ GPa  &   \\  \hline
$\phi_H = 3.76 \times 10^{-4}$ & \\  \hline
$\phi_L = 5.00 \times 10^{-5}$ & \\  \hline
$\gamma = 9.5$ & \\  \hline
\hline
\end{tabular}
\caption{Physical 
 parameters of coating and substrate material used in the simulations, 
we assume that the temperature is $T=300$ K and free-space wavelength is $\lambda_0=1064$ nm.
The two materials described above are those that have the best characteristics, mainly in terms of negligible extinction coefficients,
 and are used in both the Virgo
~\cite{Virgo} and LIGO~\cite{LIGO} experiments.
}
\label{tab:partab}
\end{table}
In Figure~\ref{fig2}a the Pareto front ${\mathcal P}_H(\bar{\phi}_c)$ of the problem (\ref{optherp}) where $N_D \le 15$ is shown in red on a log-linear 
scale {as a function of the dimensionless quantity} $\bar{\phi}_c$. 
The continuous black curve represents the Pareto front again for problem (\ref{optherp}) but with $N_D = 10$ kept fixed. 
Finally, for~fixed $N_D = 10$ the dashed curves show the Pareto front of the optimal periodic doublet design.
0

In Figure~\ref{fig2}b, a~close-up of the central area of the Figure~\ref{fig2}a  is shown along with several Pareto fronts of Herpin (continuous curves) and periodic (dashed curves) designs.

From the analysis of Figure~\ref{fig2}, 
it is clear that Herpin's design generates a Pareto front that consists of several bumps (a bumpy curve).
This behavior had already been observed in the {\it brute force} solution given by the BorgMOEA method used in the paper~\cite{OptMat}.
This figure reveals that the various bumps of the complete Pareto front are parts of the Pareto curves with fixed $N_D$ that intersect each other (see Figure~\ref{fig2}b).

\begin{figure}[b]
 \includegraphics[width=0.76\linewidth]{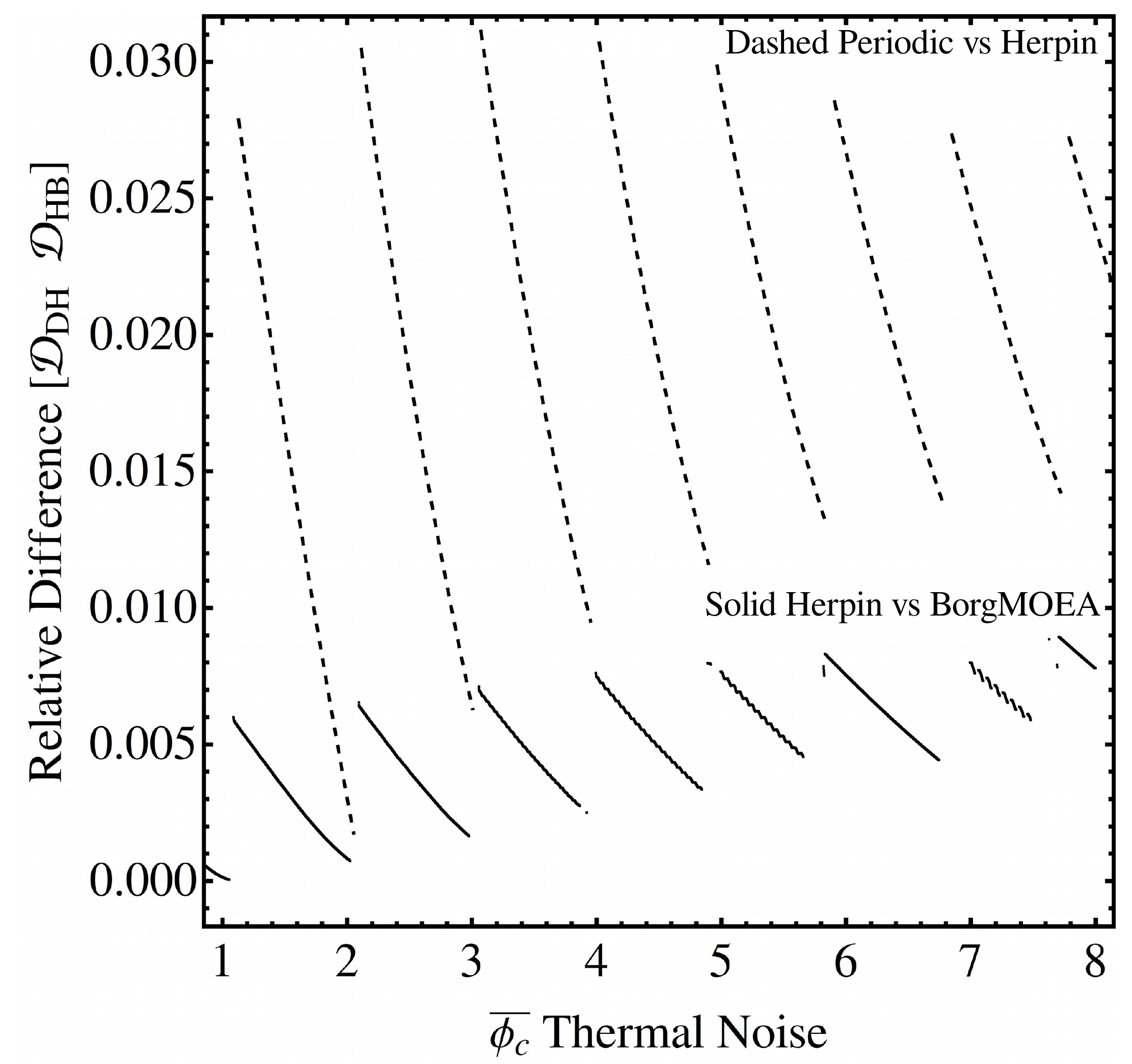}
\caption{In the figure the relative differences ${\cal D}_{DH}$ (dashed) and ${\cal D}_{HB}$ (solid) are displayed as a function of {the dimensionless} normalized mechanical loss $\bar{\phi}_c$.
The performances of Herpin design exceed the performance of periodic design and are very close to the performances of  {\it brute force} BorgMOEA numerical optimization.}
\label{fig3}
\end{figure}

Moreover, as~will be more evident later, we note that the periodic designs are always worse than Herpin's.
Let us take a closer look at this result in Figure~\ref{fig3} to better illustrate it. 
The Pareto front of the periodic synthesis of the doublet (respectively of the BorgMoea) will be called with ${\mathcal P}_D(\bar{\phi}_c)$ 
(respectively with ${\mathcal P}_B(\bar{\phi}_c)$).

In Figure~\ref{fig3} the following normalized differences 
\[
{\cal D}_{DH}={\mathcal P}_H(\bar{\phi}_c)^{-1}
[{\mathcal P}_D(\bar{\phi}_c)-{\mathcal P}_H(\bar{\phi}_c)],
\]
\beq
{\cal D}_{HB}={\mathcal P}_H(\bar{\phi}_c)^{-1}
[{\mathcal P}_H(\bar{\phi}_c)-{\mathcal P}_B(\bar{\phi}_c)],
\label{eq:nordif}
\eeq
are displayed.
To be precise, in Figure~\ref{fig3} the function ${\cal D}_{DH}$ is the dashed curve while ${\cal D}_{HB}$ is the continuous black one.
By inspection of the figure, it is clear that both the two normalized differences are {\it positive} and have
 discontinuity points in the cusps that separate the different bumps in the Pareto front curves.
The value of these functions remains limited (about 3\% for ${\cal D}_{DH}$ and about 0.9\% for ${\cal D}_{HB}$) even in the region with the highest noise 
(i.e., low transmittance), which is that of interest for gravitational~applications. 

\vspace{-3pt}

In this connection, the values of the normalized thermal noise (on the Pareto front) for the three analyzed methods are reported in Table~\ref{tab:lownoise}
for fixed values of the transmittance constraint.
Some values of the transmittance constraint are shown in the first column of  Table~\ref{tab:lownoise}
together with the value of  $N_D$ giving the optimal solutions for all the three analyzed design methods, i.e.,~periodic doublets, 
{\it brute force} BorgMOEA and Herpin design. The~values of the computed minimum normalized thermal noise $\bar{\phi}_c$ are reported in the other columns of the~Table.

This table confirms the results of Figure~\ref{fig3},  i.e.,~even in the zone of very low transmittance, the~Herpin-like semi-analytic method is the one that comes closest to the minimum value obtained with {\it brute force}, for~which there is no simple constructive~recipe.

\begin{table}[t]
\centering
\begin{tabular}{|l|l|l|l|}
\hline
\hline
\tiny \, &     \,   \tiny      &  \,  \tiny      &   \, \tiny             \\
Parameters \, &     \,   $\bar{\phi}_c$      &  \,  $\bar{\phi}_c$      &   \, $\bar{\phi}_c$             \\
   &  \small Periodic doublet &  \small Borg MOEA & \small Herpin \\ \hline \hline
\begin{tabular}[c]{@{}l@{}}$N_D=21$ \\ $\tau_c=6\, \times 10^{-6}$ \end{tabular} & 19.597   & 19.560 & 19.573  \\ \hline \hline
\begin{tabular}[c]{@{}l@{}}$N_D=20$ \\ $\tau_c=1\, \times 10^{-5}$ \end{tabular} & 18.826   & 18.786 & 18.803  \\ \hline \hline
\begin{tabular}[c]{@{}l@{}}$N_D=17$ \\ $\tau_c=6\, \times 10^{-5}$ \end{tabular} & 16.115   & 16.076 & 16.091  \\ \hline \hline
\begin{tabular}[c]{@{}l@{}}$N_D=17$ \\ $\tau_c=1\, \times 10^{-4}$ \end{tabular} & 15.345   & 15.300 & 15.317  \\ \hline \hline
\end{tabular}
\caption{The table shows the normalized mechanical loss $\bar{\phi}_c$ for fixed transmittance for three different designs. 
The target transmittance $\tau_c$ and the number of layers $N_D$  are
reported in the first column. For~these parameters, the~minimal loss angles, obtained with the doublet, BorgMOEA, and Herpin procedures are displayed in the other columns.
}
\label{tab:lownoise}
\end{table}

\section{Conclusions}
{ The production and characterization of layered systems with dimensions of hundreds of nanometers or less, to~be used as highly reflective  surfaces~\cite{bobba,refer} is a problem of great interest for improving the operation of gravitational wave antennas~\cite{Virgo, LIGO}.}
Herpin's theorem allows obtaining an equivalent stratified material, consisting of three layers arranged in a {\it palindrome} sequence, which mimic exactly a quarter-wave layer.
In this paper, these quarter-wave equivalent layers are used in conjunction with normal (quarter-wave) layers made of low refractive index material,  to~produce optimized designs of coatings.
The method reduces to an optimization problem with two independent parameters, namely the number of equivalent layers $N_D$ and the normalized thickness 
of one of the materials defining the equivalent layer. 
Thus, a~{\it turbo} solution to the problem (i.e., an~explicit analytic constructive solution) can be found in a very simple way.
Such a solution is closer to that obtained with the BorgMOEA method~\cite{OptMat} than the doublet periodic one~\cite{Spie}.
The prediction made in a previous article~\cite{OptMat}, namely that the BorgMOEA {\it brute force} design should be close to a  virtual quarter-wave design, is fully confirmed.
Indeed, in~this paper, an~explicit semi-analytic quarter-wave design is found, even of simple physical interpretation.
{ The limitation of the present study is that it only deals with the case of  binary coatings.}
The authors are convinced that with a similar philosophy it is possible to derive optimal coating designs even in the case of layers made with three (possibly dissipative) materials, or~even made with nano-layered~\cite{nanolayer} materials.

\vspace{6pt}

\section*{Acknowledgments}
This work has been partially supported by INFN
through the projects Virgo and Virgo---ET.
The author is grateful for the discussion and suggestions received from
the Virgo Coating Research and Development Group and the Optics Working
Group of the LIGO Scientific Collaboration. Special thanks to  I.M. Pinto for his constant interest and encouragement in publishing this~article.







\end{document}